\documentclass{article}
\usepackage[english]{babel}
\usepackage{amsmath,amsthm,eepic,color}
\usepackage{amsfonts}
\usepackage{graphicx}
\usepackage{bm}
\usepackage{amsmath,amsthm}
\usepackage{amssymb}
\usepackage{amsfonts,color}

\numberwithin{equation}{section}
\newtheorem{theorem}{Theorem}
\newtheorem{remark}{Remark}
\def\bbblb#1{\setbox\@tempboxa\hbox{$#1[$}%
             \@tempdimb\wd\@tempboxa
             \copy\@tempboxa \kern -.85\@tempdimb
             \copy\@tempboxa \kern -.65\@tempdimb\box\@tempboxa}
\def\bbbrb#1{\setbox\@tempboxa\hbox{$#1]$}%
             \@tempdimb\wd\@tempboxa
             \copy\@tempboxa \kern -.65\@tempdimb
             \copy\@tempboxa \kern -.85\@tempdimb\box\@tempboxa}

\def\m{{\boldsymbol m}}

\def\P{{\boldsymbol P}}

\def\n{{\bf n}}

\def\bbbc{{\Bbb C}}

\def\ad{\mbox{ad}\,}
\def\diag{\mbox{diag}\,}
\def\const{\mbox{const}\,}

\def\openone{\leavevmode\hbox{\small1\kern-3.3pt\normalsize1}}

\def\ad{\mbox{ad\,}}

\def\diag{\mbox{diag\,}}

\def\bbbc{{\Bbb C}}

\def\openone{\leavevmode\hbox{\small1\kern-3.3pt\normalsize1}}

\def\m{{\boldsymbol m}}

\newenvironment{dedication}
        {\vspace{3ex}\begin{quotation} \begin{flushright} \indent    \begin{em}}
        {\par\end{em}\end{flushright}\end{quotation}}

\def\openone{\leavevmode\hbox{\small1\kern-3.3pt\normalsize1}}

\def\diag{\mbox{diag\,}}

\def\const{\mbox{const\,}}
\def\ad{\mbox{ad\,}}

\title{On soliton solutions and soliton interactions of Kulish-Sklyanin and Hirota-Ohta systems}

\author{ V. S. Gerdjikov$^{1,6}$, Nianhua  Li$^2$, V. B. Matveev$^{3,4}$, A. O. Smirnov$^5$  \\[5pt]
{\sl $^1$Institute of Mathematics and Informatics,  Bulgarian Academy of Sciences, }\\
{\sl Acad. Georgi Bonchev Str., Block 8, 1113 Sofia, Bulgaria}\\[5pt]
{\sl $^2$School of Mathematical Sciences, Huaqiao University,}\\
{\sl  Quanzhou, Fujian 362021, P.R. China} \\[5pt]
{\sl  $^3$Institut de Math\'ematiques de Bourgogne (IMB),}\\
{\sl Universit\'e de Bourgogne - France Comt\'e, Dijon, France}\\[5pt]
{\sl $^4$Sankt Petersburg department of Steklov Mathematical} \\
{\sl  Institute of Russian Academy of Sciences}\\[5pt]
{\sl $^5$Sankt-Petersburg State  University    of   Aerospace Instrumentation }\\
{\sl St-Petersburg,   B.Morskaya,   67A,  St-Petersburg, 190000, Russia} \\[5pt]
{\sl $^6$Institute for Advanced Physical Studies,}\\
{\sl 111 Tsarigradsko chaussee, Sofia 1784, Bulgaria} }

\date{ }

\begin{document}
\maketitle
\thispagestyle{empty}

\begin{dedication}
\begin{minipage}{4.3in} {\sl
It is a great honor and pleasure for us, the authors, to dedicate this paper to Professor A. K. Pogrebkov
 with whom we have enjoyed several decades of friendship and fruitful research collaboration.
Andrey is well known as a brilliant scientist with remarkable contributions in the field of Mathematical Physics and, in particular, in the area of Integrable Systems.  For many years he served as one of the executive editors of   ''Theoretical and Mathematical Physics'', one of the leading journals in the area, and as an organizer of multiple international conferences and scientific events. We wish him a long life full of happiness and creativity.}
\end{minipage}
\end{dedication}

\begin{abstract}
In this paper we consider a simplest two-dimensional reduction of the remarkable three-dimensional Hirota-Ohta system. The Lax pair of the Hirota-Ohta system was extended to a Lax triad by adding extra third linear equation, whose compatibility conditions with the Lax pair of the Hirota-Ohta imply another remarkable systems: the Kulish-Sklyanin system (KSS) together with its first higher commuting flow, which we can call as vector complex MKdV.
This means that any common particular solution of these both two-dimensional integrable systems yields a corres\-ponding particular solution of the three-dimensional Hirota-Ohta system. Using the dressing Zakharov-Shabat method we derive the $N$-soliton solutions of these systems and analyze their interactions, i.e. derive expli\-citly the shifts of the relative center-of-mass coordinates and the phases as functions of the discrete eigenvalues of the Lax operator. Next we relate to these nonlinear evolution equations (NLEE) system of Hirota--Ohta type and obtain its $N$-soliton solutions

 \noindent{\bf Keywords:}  Two-dimensional Kulish-Sklyanin system, Three-dimensional Hirota-Ohta system, Lax representation, Dressing method, Multi-soliton solutions, Two-dimensional reductions

\end{abstract}

\pagenumbering{arabic}

\section{Introduction}
$2+1$-dimensional integrable equations play important role in today mathematics  and its applications to physics. The first such equation 
were the famous KP-I and KP-II equations having important physical applications, see \cite{ACTV, Matv, Matv1, BoiPog, BoiPog2} and references therein.
Different classes of solutions for these equations   and the related hierarchies
were obtained  using quite different approaches. The first one was  the  Zakharov-Shabat dressing method \cite{ZaSha1, ZMNP}
 applicable both  to the KP-I and KP-II case. It works for instance for the KP-I equation, allowing to construct the so-called multi-lump rational solutions 
 which are  smooth, real rational  solutions  described  by explicit determinant formulas \cite{MaZaBoItMat}.
 Much larger family of non-singular solutions was described in \cite{ACTV} by means of the Grammian determinant formula, developed much earlier
 in \cite{MatSalle} and \cite{MatSalle2}; for more details see, e.g. \cite{EPJPI, EPJPII}. Latest results on the subject can be traced, e.g. from 
 \cite{BoiPog3, Pog} and the references therein. 


Our aim with this paper is first to reformulate some of our results on soliton solutions of $1+1$ dimensional multi-component NLS-type equations for somewhat more general context. Next, these more general results will be used to derive soliton solutions of certain $2+1$-dimensional equations.

Indeed, it is often the case that now many authors are able to solve the system \cite{ZMNP, FaTa, Mon1, DokLeb}:
\begin{equation}\label{eq:1.1}\begin{split}
 i q_t + q_{xx} + 2 q^2 p(x,t) &= 0, \\
 -i p_t + p_{xx} + 2 p^2 q(x,t) &= 0.
\end{split}\end{equation}
Instead most of them quickly specify that they are going to solve the important particular case:
\begin{equation}\label{eq:nls}\begin{split}
  i q_t + q_{xx} + 2 \epsilon |q|^2 q(x,t) &= 0,
\end{split}\end{equation}
discovered by \cite{ZaSha}. Eq. (\ref{eq:nls}) is obtained from (\ref{eq:1.1}) after the involution $p=\epsilon q^*$, $\epsilon =\pm 1$. There is an obvious reason for this. The nonlinear Schr\"odinger (NLS) equation  (\ref{eq:nls}) finds numerous applications in nonlinear optics, plasma physics, hydrodynamics etc. At the same time the system (\ref{eq:1.1}) to the best of our knowledge does not allow such variety of applications.

In the first part of our paper we will analyze two multicomponent systems generalizing  (\ref{eq:1.1}), namely:
\begin{equation}\label{eq:ks2}\begin{split}
 i \vec{q}_{t} + \vec{q}_{xx} + 2 (\vec{q}^T. \vec{p}) \vec{q}  -(\vec{q}^T s_0 \vec{q}) \vec{p}(x,t) &=0, \\
 -i \vec{p}_{t} + \vec{p}_{xx} + 2 (\vec{q}^T. \vec{p}) \vec{p}  -(\vec{p}^T s_0 \vec{p}) \vec{q}(x,t) &=0,
\end{split}\end{equation}
and
\begin{equation}\label{eq:ks3}\begin{split}
\vec{q}_{y} + \vec{q}_{xxx} + 3 (\vec{q}_x^T.\vec{p}) \vec{q} + 3 (\vec{q}^T. \vec{p}) \vec{q}_x - 3(\vec{q}^T s_0 \vec{q}_x) s_0\vec{p}(x,t) &=0, \\
\vec{p}_{y} + \vec{p}_{xxx} + 3 (\vec{p}_x^T.\vec{q}) \vec{p} + 3 (\vec{q}^T. \vec{p}) \vec{p}_x - 3(\vec{p}^T s_0 \vec{p}_x) s_0\vec{q}(x,t) &=0.
\end{split}\end{equation}
see \cite{MatSmi2}. Here $\vec{q}(x,t)$ and $\vec{p}(x,t)$ (or $\vec{q}(x,y)$ and $\vec{p}(x,y)$) are $2n-1$-component complex-valued vector functions of $x$ and $t$ (or $x$ and $y$) and
\begin{equation}\label{eq:s0}\begin{split}
 s_0 = \sum_{k=1}^{2n-1} (-1)^{2n-k} E_{k,2n-k},
\end{split}\end{equation}
Equation (\ref{eq:ks2}) with $\vec{p} = \vec{q}\;^*$ has been discovered by Kulish and Sklyanin \cite{KuSkl} about three decades ago. Below we will call it and its more general form (\ref{eq:ks2}) Kulish-Sklyanin system (KSS).

Both systems (\ref{eq:ks2}) and (\ref{eq:ks3}) allow Lax representation with the same Lax operator:
\begin{equation}\label{eq:Lax}\begin{split}
 L \psi \equiv \left( \frac{\partial \psi}{ \partial x } + Q(x,t) - \lambda J \right) \psi(x,t,\lambda) = 0, \qquad Q(x,t) = \left(\begin{array}{ccc} 0 & \vec{q}^T & 0 \\ \vec{p} & 0 & s_0 \vec{q} \\ 0 & \vec{p}^T s_0 & 0   \end{array}\right).
\end{split}\end{equation}
Here $Q(x,t)$ is $(2n+1)\times (2n+1)$ matrix-valued function  and $ J = \diag ( 1, 0,\dots, 0, -1)$. Both $Q(x,t)$ and $J$ are elements of the Lie algebra $so(2n+1)$. Here and below we will say that $Q$ is an element of the orthogonal algebra $so(2n+1)$ if
\begin{equation}\label{eq:QS0}\begin{split}
 Q + S_0 Q^T S_0 = 0, \qquad  S_0 = \sum_{k=1}^{2n+1} (-1)^{2n+2-k} E_{k,2n+2-k}, \qquad (E_{sr})_{jk} = \delta_{sj} \delta_{rk} .
\end{split}\end{equation}
This choice is convenient because the Cartan subalgebra elements (one of them is $J$) are given by diagonal matrices. Readers familiar with Lie algebras and symmetric spaces will recognize $Q(x,t)$ as the local coordinate of the symmetric space of BD.I type $SO(2n+1)/(SO(2n-1)\times SO(2))$ \cite{Helg}. The deep relation between the symmetric spaces and the multicomponent NLS equations was discovered by Kulish and Fordy \cite{ForKu*83}.

Indeed, the KSS is the compatibility condition $[L,M_2]=0$  between  $L$ and $M_2$:
\begin{equation}\label{eq:M2}\begin{split}
M_2\psi (x,t,\lambda ) &\equiv  i\partial_t\psi + (V_1^0(x,t) + \lambda Q(x,t) - \lambda ^2 J)\psi  (x,t,\lambda )=0, \\
V_1^0(x,t) &= i \ad_J^{-1} \frac{d Q}{dx} + \frac{1}{2} \left[\ad_J^{-1} Q, Q(x,t) \right], \qquad J=\diag (1,0,\ldots 0, -1).
\end{split}\end{equation}

The equation (\ref{eq:ks3}) also possesses Lax representation $[L,M_3]=0$ with the same type of Lax operator $L$, only now $Q = Q(x,y)$ is function of $x$ and $y$   and \cite{SGA}:
\begin{equation}\label{eq:M3}\begin{split}
M_3\psi (x,y,\lambda )
&\equiv  i\frac{\partial }{\partial y}\psi + (V_2^0(x,y) + \lambda V_1^0(x,y) + \lambda^2 Q(x,y) - \lambda ^3 J)\psi  (x,y,\lambda )=0, \\
V_2^0(x,y) &= i \ad_J^{-1} \frac{\partial^2 Q}{\partial x^2} + \frac{i}{2}\ad_J^{-1}  \left[\ad_J^{-1} Q(x,y), \frac{\partial Q}{ \partial x}  \right].
\end{split}\end{equation}

In what follows from the context it will be clear whether we have potential $Q(x,t)$, $Q(x,y)$ or $Q(x,t,y)$.

\begin{remark}\label{rem:0}
In the generic case when $x$, $t$ and $y$ are three independent variables we will have three operators which commute between themselves. Therefore they will have common system of fundamental analytic solutions (FAS) $\chi^\pm (x,t,y,\lambda)$. Below we will have three options: i) consider the hierarchy related to $L$ and $M_2$ with $Q(x,t)$; ii) consider the hierarchy related to $L$ and $M_3$ with $Q(x,y)$; and iii) consider the hierarchy related to the triple $L$, $M_2$ and $M_3$ with $Q(x,t,y)$.

Obviously, if $y \propto t$ then both equations (\ref{eq:ks2}) and (\ref{eq:ks3}) belong to the hierarchy of nonlinear evolution equations (NLEE).
\end{remark}

Both systems (\ref{eq:ks2}) and (\ref{eq:ks3}) allow involutions $\vec{p} = \vec{q}^*$. After that the system (\ref{eq:ks2}) goes into
\begin{equation}\label{eq:ks20}\begin{split}
 i \vec{q}_{t} + \vec{q}_{xx} + 2 (\vec{q}^\dag. \vec{q}) \vec{q}  -(\vec{q}^T s_0 \vec{q}) \vec{q}^*(x,t) &=0,
\end{split}\end{equation}
and equation (\ref{eq:ks3}) takes the form:
\begin{equation}\label{eq:ks30}\begin{split}
\vec{q}_{y} + \vec{q}_{xxx} + 3 (\vec{q}_x^\dag.\vec{p}) \vec{q} + 3 (\vec{q}^\dag. \vec{q}) \vec{q}_x - 3(\vec{q}^T s_0 \vec{q}_x) s_0\vec{q}^*(x,t) &=0,
\end{split}\end{equation}
The system (\ref{eq:ks3}) allows additional involutions. Besides the typical one $\vec{p} =\vec{q}\;^*$  they allow $\vec{q} =\vec{q}^*$ and  $\vec{p} =\vec{p}^*$, as well as  $\vec{q} =\vec{p}$ for the real-valued vectors  $\vec{q}$ and $\vec{p}$.

One of its important applications is that the three-component KSS ($n=2$)  describes one-dimensional spin-1 Bose-Einstein condensate, see \cite{IMW04,86, AMITANS, Prinari}. Likewise, five-component ($n=3$) KSS describes one-dimensional Bose-Einstein condensate (BEC) with spin 2.
The construction of the fundamental analytic solutions (FAS) of the Lax operator (\ref{eq:Lax}) was done in \cite{86, vgn2a}. The derivation of the soliton solutions may be viewed as a part of a more general theory that can be applied to the more general class of multicomponent nonlinear Schr\"odinger (MNLS) equations \cite{MaZa*76, ForKu*83, Basic, 142a, GGK05a}. Such MNLS equations can be related to each of the known Hermitian symmetric spaces \cite{ForKu*83}. The fundamental result of AKNS \cite{AKNS} concerning the interpretation of the inverse scattering method (ISM) as a generalized Fourier transform was extended also for the MNLS equations \cite{ VSG2, Holyoke, Mon1, EPJPI, EPJPII, G*86}.

The KSS belongs to a hierarchy of integrable NLEE related to the same Lax operator (\ref{eq:Lax}).  The next equations in this hierarchy (\ref{eq:ks3}), as mentioned above, allow additional involutions. All such systems of  NLEE are integrable and possess compatible hierarchies of Hamiltonian structures \cite{Holyoke, Basic,  EPJPI, EPJPII}.

This paper extends previous results belonging to us and our collaborators \cite{PLA126, VSGG*10a, VSGG*10b, 103, 142a, 86, MatSmi, MatSmi2, Smi3}.

In the next Section 2 we collect preliminary results about the direct and inverse scattering theory of the Lax operator $L$ (\ref{eq:Lax}). In Section 3 we derive the soliton solutions of (\ref{eq:ks2}) and (\ref{eq:ks3}) and analyze their interactions. The results there are generalizations of our earlier results \cite{142a, vgrn, GGK05a}.
In Section 4 we establish the interrelation between (\ref{eq:ks2})  and (\ref{eq:ks3}) and the Hirota-Ohta system \cite{HiOh1, HiOh2, Adler}. In particular knowing the soliton solutions of (\ref{eq:ks2})  and (\ref{eq:ks3}) we are able to derive the soliton solutions of Hirota-Ohta.

\section{ Preliminaries}
The inverse scattering problem for the Lax operator $L$ can be reduced to a Riemann-Hilbert problem, see \cite{ZMNP, ZaMi} for the general case and \cite{AMITANS, 86, Rosen} for the {\bf BD.I}-type MNLS. Here we will consider the general case iii) of Remark \ref{rem:0}. The cases i) and ii) can easily be recovered from iii) neglecting the $t$ or $y$ dependence in the sewing function.

\subsection{Direct Scattering Problem for $L$}\label{ssec:2.1}

The starting point for solving the direct and the inverse scattering problem (ISP) for
$L$ are  the so-called Jost solutions,  which are defined by their asymptotics (see, e.g. \cite{VSGG*10a, VSGG*10b}
and the references therein):
\begin{equation}\label{eq:Jost}
\lim_{x \to -\infty} \phi(x,t,y,\lambda) e^{  i \lambda J x }=\openone, \qquad  \lim_{x \to \infty}\psi(x,t,y,\lambda) e^{  i\lambda J x } = \openone,
 \end{equation}
and the scattering matrix $T(\lambda,t,y)$ is defined by $T(\lambda,t,y)\equiv
\psi^{-1}\phi(x,t,y,\lambda)$. Here we assume that the potential $Q(x,t,y)$ is tending to zero fast enough, when $|x|\to \infty$. The special choice of $J$ results in the
fact that the Jost solutions and the scattering matrix take values
in the corresponding orthogonal Lie group $SO(2n+1)$. One can use the following block-matrix
structure of $T(\lambda,t,y)$
\begin{equation}\label{eq:25.1}
T(\lambda,t,y) = \left( \begin{array}{ccc} m_1^+ & -\vec{b}^-{}^T & c_1^- \\
\vec{b}^+ & {\bf T}_{22} & - s_0\vec{B}^- \\ c_1^+ & \vec{B}^+{}^Ts_0 & m_1^- \\
\end{array}\right), \qquad \hat{T}(\lambda,t,y) = \left( \begin{array}{ccc} m_1^- & \vec{B}^-{}^T & c_1^- \\
-\vec{B}^+ & {\bf \hat{T}}_{22} &  s_0\vec{b}^- \\ c_1^+ & -\vec{b}^+{}^Ts_0 & m_1^+ \\
\end{array}\right),
\end{equation}
where $\vec{b}^\pm (\lambda,t,y)$ and $\vec{B}^\pm (\lambda,t,y)$ are $2n-1$-component vectors, ${\bf T}_{22}(\lambda)$ and ${\bf \hat{T}}_{22} (\lambda)$ are $(2n-1) \times (2n-1)$ block matrices, and $m_1^\pm (\lambda)$, $c_1^\pm (\lambda)$ are scalar functions. Here  and below by `hat' we will denote taking the inverse, i.e. $\hat{T}(\lambda, t,y) =T^{-1}(\lambda,t,y)$. Such parametrization is compatible with the generalized Gauss decompositions \cite{Helg} of $T(\lambda,t,y)$.

With this notations we introduce the generalized Gauss factors of
$T(\lambda)$ as follows:
\begin{equation}\label{eq:25.1'}
T(\lambda,t,y) = T^-_J D^+_J \hat{S}^+_J  = T^+_J D^-_J \hat{S}^-_J ,
\end{equation}

\begin{equation}\label{eq:25.1''}
\begin{aligned}
T^-_J(\lambda, t, y)  &=\left( \begin{array}{ccc}  1 & 0 & 0 \\ \vec{\rho}^+ & \openone & 0 \\
c_1^{\prime\prime,+} & \vec{\rho}^{+,T}s_0 & 1 \\ \end{array} \right), &\quad
T^+_J(\lambda, t, y)  &= \left( \begin{array}{ccc}  1 & -\vec{\rho}^{-,T} & c_1^{\prime\prime,-} \\ 0
& \openone & -s_0\vec{\rho}^- \\ 0 & 0 & 1 \\ \end{array} \right),\\
 S^+_J(\lambda, t, y) &= \left(\begin{array}{ccc}  1 & \vec{\tau}^{+,T} & c_1^{\prime,-} \\ 0 & \openone & s_0\vec{\tau}^+ \\ 0 & 0 & 1 \\ \end{array} \right),  &\quad S^-_J (\lambda, t, y) &=
\left( \begin{array}{ccc}  1 & 0 & 0 \\ -\vec{\tau}^- & \openone & 0 \\
c_1^{\prime,+} & -\vec{\tau}^{-,T}s_0 & 1 \\ \end{array} \right), \\
D_J^+ (\lambda) &= \left( \begin{array}{ccc} m_1^+ & 0 & 0 \\ 0 & \m_2^+ & 0 \\
0 & 0 & 1/m_1^+ \end{array} \right), &\quad  D_J^-(\lambda) &=
\left( \begin{array}{ccc} 1/m_1^- & 0 & 0 \\ 0 &  \m_2^- & 0 \\
0 & 0 & m_1^- \end{array} \right),
\end{aligned}\end{equation}
where
\begin{equation}\label{eq:25.1a}
\begin{aligned}
c_1^{\prime\prime,\pm} &= \frac{m_1^\pm}{2}(\vec{\rho}^{\pm,T} s_0 \vec{\rho}^\pm) , &\quad c_1^{\prime,\pm} &=
\frac{m_1^\mp}{2}(\vec{\tau}^{\mp,T} s_0 \vec{\tau}^\mp), \\
\vec{\rho}^-(\lambda, t, y) &=\frac{\vec{B}^-}{m_1^-}, \quad \vec{\tau}^- (\lambda, t, y) =\frac{\vec{B}^+}{m_1^-}, &\quad \vec{\rho}^+(\lambda, t, y) &=\frac{\vec{b}^+}{m_1^+}, \quad \vec{\tau}^+ (\lambda, t, y) =\frac{\vec{b}^-}{m_1^+},\\
\m_2^+(\lambda) &= {\bf T}_{22} + \frac{\vec{b}^+ \vec{b}^-{}^T  }{m_1^+}, &\quad \m_2^-(\lambda)  &= {\bf T}_{22} +  \frac{s_0\vec{b}^- \vec{b}^+{}^T s_0 }{m_1^-}.
\end{aligned}
\end{equation}

Important tools for reducing the ISP to a Riemann-Hilbert problem (RHP) are the fundamental analytic solution (FAS) $\chi^{\pm} (x,t,\lambda )$. Their construction is based on the generalized Gauss decomposition of $T(\lambda,t)$, see \cite{ZMNP, Basic, TMF98}:
\begin{equation}\label{eq:FAS_J}
\chi ^\pm(x,t,y,\lambda)= \phi (x,t,y,\lambda) S_{J}^{\pm}(t,y,\lambda )
= \psi (x,t,y,\lambda ) T_{J}^{\mp}(t,y,\lambda ) D_J^\pm (\lambda).
\end{equation}

The two analyticity regions $\bbbc_+$ and $\bbbc_-$ are separated by the real line. The continuous spectrum of $L$ fills
in the real line and has multiplicity $2$, see Subsection 3.3 below.

The Lax representation ensures that if $Q(x,t,y) $ evolves according to (\ref{eq:ks2}) and (\ref{eq:ks3}) then the scattering matrix and its elements satisfy the following linear evolution equations
\begin{equation}\label{eq:evks2}
\begin{aligned}
 i\frac{d\vec{b}^{\pm}}{d t} \pm \lambda ^2 \vec{b}^{\pm}(t,y, \lambda ) &=0, &\qquad i\frac{d\vec{B}^{\pm}}{d t} \pm \lambda ^2\vec{B}^{\pm}(t,y,\lambda ) &=0, \\
 i\frac{d\vec{b}^{\pm}}{d y} \pm \lambda ^3 \vec{b}^{\pm}(t,y,\lambda ) &=0, &\qquad i\frac{d\vec{B}^{\pm}}{d y} \pm \lambda ^3 \vec{B}^{\pm}(t,y,\lambda ) &=0, \\
 i\frac{d m_1^{\pm}}{d t}  &=0, &\qquad  i \frac{d \m_2^{\pm}}{d t}  &=0,
\end{aligned}
\end{equation}

The block-diagonal matrices $D^{\pm}(\lambda)$ are time-independent functions of $\lambda$. Thus they can be considered as generating functionals of the integrals of motion.
It is well known \cite{DrSok, ForKu*83, Basic} that generic nonlinear evolution equations related to a simple Lie algebra $\mathfrak{g} $ of rank $n$ possess $n$ series of integrals of motion in involution; thus for them one can prove complete integrability \cite{BeSat}. In our case, beside $m_1^\pm(\lambda)$, we can consider as generating functionals of integrals of motion  all $(2n-1)^2$ matrix elements of $\m_2^\pm(\lambda)$, for $\lambda \in \bbbc_\pm$.
However they can not be all in involution. Such situation is characteristic for the superintegrable models. It is due to the degeneracy of the dispersion law in (\ref{eq:evks2}). We remind that $D^\pm_J(\lambda)$ allow analytic extension for $\lambda\in \bbbc_\pm$ and that their zeroes and poles determine the discrete eigenvalues of $L$.

\subsection{Equivalence of the Lax representation to a Riemann-Hilbert Problem }\label{ssec:2.2}

The FAS for real $\lambda$ are linearly related
\begin{equation}\label{eq:rhp0}
\chi^+(x,t,y,\lambda) =\chi^-(x,t,y,\lambda) G_{0,J}(\lambda,t,y),
\qquad G_{0,J}(\lambda,t,y)=\hat{S}^-_J(\lambda,t,y)S^+_J(\lambda,t,y) .
\end{equation}
One can rewrite equation (\ref{eq:rhp0}) in an equivalent form for the FAS
\begin{equation}\label{eq:xipm}\begin{split}
\xi^\pm(x,t,y,\lambda)=\chi^\pm (x,t,y,\lambda)e^{i\lambda Jx }
\end{split}\end{equation}
which satisfy the equation:
\begin{equation}\label{eq:xi}
i\frac{d\xi^\pm}{dx} + Q(x)\xi^\pm(x,t,y\lambda) -\lambda [J,
\xi^\pm(x,t,y,\lambda)]=0, \qquad \lim_{\lambda \to \infty} \xi^\pm(x,t,y,\lambda) = \openone.
\end{equation}
Depending on the choice of the second operator $M_2$ or $M_3$, $\xi^\pm(x,t,\lambda)$ will satisfy also
\begin{equation}\label{eq:xit2}
i\frac{d\xi^\pm}{dt} + (V_1^0(x,t,y) + \lambda Q(x,t,y))\xi^\pm(x,t,y,\lambda) -\lambda^2 [J, \xi^\pm(x,t,y,\lambda)] =0,
\end{equation}
and
\begin{equation}\label{eq:xit3}
i\frac{d\xi^\pm}{dy} + (V_2^0(x,t,y) + \lambda V_1^0(x,t,y) + \lambda^2 Q(x,t,y))\xi^\pm(x,t,y,\lambda) -\lambda^3 [J, \xi^\pm(x,t,y,\lambda)] =0.
\end{equation}
These two FAS must also be linearly related:
\begin{equation}\label{eq:rhp1}
\xi^+(x,t,y,\lambda) =\xi^-(x,t,y,\lambda) G_J(x,t,y,\lambda), \qquad
G_{J}(x,t,y,\lambda) =e^{-i\lambda Jx}G^-_{0,J}(\lambda,t,y)e^{i\lambda Jx} .
\end{equation}
Obviously the sewing function $G_J(x,t,y, \lambda)$ is uniquely determined by the Gauss factors $S_J^\pm (\lambda,t,y)$. Equation (\ref{eq:rhp1}) is a Riemann-Hilbert problem (RHP) in multiplicative form.  As mentioned above in eq. (\ref{eq:xi}) they also satisfy
\begin{equation}\label{eq:xi8}\begin{split}
 \lim_{\lambda \to \infty} \xi^\pm(x,t,y,\lambda)  = \openone,
\end{split}\end{equation}
which is known as the canonical normalization of the RHP (\ref{eq:rhp1}).

Thus we demonstrated that the FAS of the Lax triple satisfy the RHP (\ref{eq:rhp1}) with canonical normalization (\ref{eq:xi8}).

The next important step is based on the theorem below which states, that the inverse is also true.
\begin{theorem}[Zakharov-Shabat,  \cite{ZaSha1, ZaSha2}]\label{thm:ZaSha}
 Let $\xi^+(x,t,y,\lambda)$ and $\xi^-(x,t,y,\lambda)$ be solutions to the RHP (\ref{eq:rhp1}) with canonical normalization (\ref{eq:xi8}), and let the sewing function satisfy:
 \begin{equation}\label{eq:dGJxt2}\begin{split}
  i \frac{\partial G_J}{ \partial x } - \lambda [J, G_J(x,t,y,\lambda)]=0, \qquad
  i \frac{\partial G_J}{ \partial t } - \lambda^2 [J, G_J(x,t,y,\lambda)]=0, \\
  i \frac{\partial G_J}{ \partial y } - \lambda^3 [J, G_J(x,t,y,\lambda)]=0.
 \end{split}\end{equation}
Then $\xi^\pm(x,t,y,\lambda)$ are FAS of the Lax triple $L$, $M_2$ and $M_3$.
\end{theorem}

\begin{remark}\label{rem:1}
The proof of this theorem is well known and we will skip it. The fact that we are dealing with generic Lax operator for which $Q(x,t,y) \neq Q^\dag (x,t,y)$ does not influence the proof. Of course one should keep in mind that the discrete eigenvalues $\lambda_j^\pm$ of $L$ correspond to singularities of the solutions of RHP. So in the generic case we may have  $\lambda_j^+ \neq  \lambda_j^{-,*}$.

\end{remark}

\section{Soliton solutions and their interactions}

\subsection{The dressing method for generic  BD.I Lax operators}

The soliton solutions can be derived by an appropriate modification \cite{86} of the  Zakharov-Shabat dressing method \cite{ZaSha1, ZaSha2}, see also \cite{VSGG*10a, VSGG*10b, 103, MatSmi, MatSmi2, Smi3}. Applying it to generic BD.I Lax operators means that, first, $L$ has the form (\ref{eq:Lax}) and, in addition, the potential $Q(x,t,y)$ is not constrained by any involution.

Let us assume that we know the regular FAS $\chi_0^\pm (x,t,y,\lambda)$ of the Lax operator $L_0$ with potential $Q_0(x,t,y)$. By regular we mean that the operator $L_0$ does not have discrete eigenvalues and as a result the  solutions $\xi_0^\pm (x,t,y,\lambda)$ of the corresponding RHP will be regular for all $\lambda \in \mathbb{C}_\pm$. Next we wish to construct new solutions $\xi^\pm (x,t,y,\lambda)$ of the RHP (\ref{eq:rhp1}) which possess simple pole singularities at the $N$ pairs of points $\lambda_j^\pm \in \mathbb{C}_\pm $. We will denote the `dressed` operator with potential $Q(x,t,y)$ by $L$. To this end we will introduce the dressing factor $u(x,t,y,\lambda)$ as follows:
\begin{equation}\label{eq:u}\begin{split}
\xi^\pm (x,t,y,\lambda) = u (x,t,y,\lambda) \xi_0^\pm (x,t,y,\lambda).
\end{split}\end{equation}
Therefore the dressing factor must have $2N$ simple poles at the points $\lambda_j^\pm$. In addition both $\xi_0^\pm (x,t,y,\lambda)$ and $\xi^\pm (x,t,y,\lambda)$ must satisfy the canonical normalization (\ref{eq:xi8}). Therefore the dressing factor must have the form:
\begin{equation}\label{eq:u1}\begin{split}
u(x,t,y,\lambda)&=\openone+\sum^N_{k=1}\left((c_k(\lambda)-1)P_k(x,t,y) + (c_k^{-1}(\lambda)-1) \bar{P}_k(x,t,y) \right), \qquad
 c_k(\lambda) = \frac{\lambda -\lambda_k^+ }{\lambda -\lambda_k^-}.
\end{split}\end{equation}

\begin{remark}\label{rem:gen}
Applying such dressing factors we will be able to construct $N$-soliton solutions of KSS and mKdV,  i.e. to each soliton  we  associate a pair of discrete eigenvalues of  $L$:
\[ \lambda_j^+ = \mu_j^+ + i \nu_j^+ \in \mathbb{C}_+, \qquad \lambda_j^- = \mu_j^+ - i \nu_j^-  \in \mathbb{C}_-, \]
where $\mu_j^\pm$ and $\nu_j^\pm >0$ are real constants.  Usually in the cases analyzed before \cite{ZaSha1, ZaSha2, Basic, EPJPI, EPJPII} the authors impose an involution on the potential like $\vec{p} =\pm \vec{q}{\;}^*$ which ensures that $\lambda_j^- = \lambda_j^{+,*}$. Here we have generic case with $\lambda_j^- \neq \lambda_j^{+,*}$, which makes the calculations slightly more involved. If in the final results we put $\mu_j^+=\mu_j^-$ and  $\nu_j^+=\nu_j^-$ we recover the results for the typical cases with involutions.
\end{remark}

The next important property of the dressing factor follows from the fact that both  $\xi_0^\pm (x,t,y,\lambda)$ and $\xi^\pm (x,t,y,\lambda)$ are elements of the Lie group $SO(2n+1)$; therefore $u(x,t,y,\lambda)$ must also be an element of $SO(2n+1)$:
\begin{equation}\label{eq:um1}\begin{split}
u^{-1}(x,t,y,\lambda) &= S_0 u^T(x,t,\lambda) S_0 \\
&= \openone+\sum^N_{k=1}\left((c_k(\lambda)-1)\bar{P}_k(x,t,y) + (c_k^{-1}(\lambda)-1) P_k(x,t,y) \right).
\end{split}\end{equation}
where $S_0$  was introduces in (\ref{eq:QS0}). Thus in order to find $u(x,t,y,\lambda)$ we need: i) to ensure that $u(x,t,y,\lambda)$ satisfies (\ref{eq:um1}) and ii) ensure that the solutions of the RHP  $\xi_0^\pm (x,t,y,\lambda)$ and $\xi^\pm (x,t,y,\lambda)$ satisfy equation (\ref{eq:xi}),  (\ref{eq:xit2}) and (\ref{eq:xit3}) with potentials $Q_0(x,t,y)$ and $Q(x,t,y)$ respectively.

The condition i) means that $u(x,t,y,\lambda) u^{-1}(x,t,y,\lambda)=\openone$, i.e. we obtain a set of algebraic relations between $P_k$ and $\bar{P}_k$. In particular the product $u(x,t,y,\lambda) u^{-1}(x,t,y,\lambda)$ for generic $P_k$ and $\bar{P}_k = S_0 P_k^T S_0$ allows second order poles at $\lambda_k^+$ and $\lambda_k^-$. Their  residues will vanish if:
\begin{equation}\label{eq:PkPk}\begin{split}
 P_k \bar{P}_k = \bar{P}_k P_k.
\end{split}\end{equation}
At this point for the sake of simplicity we will assume that $P_k$ and $\bar{P}_k$ are rank 1 projectors and admit the following decomposition \cite{ZaMi,GGK05a}:
\begin{equation}\label{eq:Pk}\begin{split}
P_k(x,t,y)= |N_k(x,t,y)\rangle \langle m_k(x,t,y)|,\qquad \bar{P}_k(x,t,y)=|M_k(x,t,y)\rangle \langle n_k(x,t,y)|.
\end{split}\end{equation}
Thus  we obtain the first constrains on  $P_k$ and $\bar{P}_k$:
\begin{equation}\label{eq:mm0}\begin{split}
\langle m_k(x,t,y)| S_0 |m_k(x,t,y) \rangle =0, &\qquad \langle n_k(x,t,y)| S_0 |n_k(x,t,y) \rangle =0, \\ \langle N_k(x,t,y)| S_0 |N_k(x,t,y) \rangle =0, &\qquad \langle M_k(x,t,y)| S_0 |M_k(x,t,y) \rangle =0.
\end{split}\end{equation}

\begin{remark}\label{rem:1b}
For $N=1$ the constraints in eq. (\ref{eq:mm0}) and the fact that $P_k^2 =P_k$ and $\bar{P}_k^2 = \bar{P}_k$ ensure that $u(x,t,y,\lambda) \in SO(2n+1)$. For $N>1$ we need additional constraints.
\end{remark}

In order to satisfy condition ii) above we rewrite the dressing factor as $u(x,t,y,\lambda) = \xi^+(x,t,y,\lambda) \hat{\xi}_0^+(x,t,y,\lambda)$ and use eq. (\ref{eq:xi}). Here by `hat` we denote the inverse matrix, i.e. $\hat{X} = X^{-1}$. Thus we obtain:
\begin{equation}\label{eq:uum1}\begin{split}
i \frac{\partial u}{ \partial x } &+(Q(x,t,y) -\lambda J) u(x,t,y,\lambda) - u(x,t,y,\lambda) (Q_0(x,t,y) -\lambda J) =0, \\
i \frac{\partial \hat{u}}{ \partial x } &+(Q_0(x,t,y) -\lambda J) \hat{u}(x,t,y,\lambda) - \hat{u}(x,t,y,\lambda) (Q(x,t,y) -\lambda J) =0.
\end{split}\end{equation}
These two equations must hold identically with respect to $\lambda$, so all the residues of their $2N$ poles must vanish. This leads to a set of algebraic equations for $P_k$ and $\bar{P}_k$.
This is one of the best known versions of the dressing method.

The last step in the dressing method consists in determining the potential of the `dressed` operator. This is done by considering the equations (\ref{eq:uum1}) for $\lambda \to \infty$ with the result:
\begin{equation}\label{eq:QNs}
\begin{split}
& Q_{\rm Ns}(x,t,y) - Q_0(x,t,y) \\
 &= \lim_{\lambda \to \infty} \lambda (J - uJu^{-1} (x,t,y,\lambda)) = \left[ J, \sum_{k=1}^N (\lambda_k^- - \lambda_k^+) (P_k  - \bar{P}_k)  \right].
\end{split}
\end{equation}
One may say, that this procedure allows one to add $N$-solitons on top of the regular potential $Q_0(x,t,y)$.

However below it will be more convenient to us to use  an alternative version of the dressing method. So let us finish this subsection taking $N=1$. In this case solving the equations coming from (\ref{eq:uum1}) we obtain the projectors $P_1$ and $\bar{P}_1$ in the form:
\begin{equation}\label{eq:P1P1}\begin{aligned}
& P_1(x,t,y)  = \frac{|n_1(x,t,y) \rangle \langle m_1(x,t,y)| }{\langle m_1(x,t,y)|n_1(x,t,y) \rangle }, &\quad  &\bar{P}_1(x,t,y) = \frac{ S_0|n_1(x,t,y) \rangle \langle m_1(x,t,y)|S_0 }{\langle m_1(x,t,y)|n_1(x,t,y) \rangle }, \\
& |n_1(x,t,y)\rangle =\chi^{+}_0(x,t,y,\lambda^+_1)|n_{01}\rangle  ,&\quad & \langle m_1(x,t,y)| = \langle m_{0k}| \hat{\chi}^{-}_0(x,t,y,\lambda^-_1).
\end{aligned}\end{equation}
where  $|n_{01}\rangle$ and $\langle m_{01}| $ are $2n+1$-component constant vectors, known as polarization vectors.

For the truly soliton solutions $Q_0(x,t,y) =0$ and therefore $\chi^{\pm}_0(x,t,y,\lambda) =e^{-i\lambda(x+\lambda t+\lambda^2 y)J}$. From eq. (\ref{eq:P1P1}) it follows  that $P_1$ and $\bar{P}_1$ will satisfy (\ref{eq:PkPk}) we choose properly the polarization vectors so that:
\begin{equation}\label{eq:m0s0m0}\begin{split}
\langle m_{k}|S_0|m_{k} \rangle =  \langle m_{0k}|S_0|m_{0k} \rangle = 0, \qquad  \langle n_{k}|S_0|n_{k} \rangle = \langle n_{0k}|S_0|n_{0k} \rangle =0.
\end{split}\end{equation}

Now we introduce proper  parametrization for $|n_{0k}\rangle $ and $\langle m_{0k}|$ as follows:
\begin{equation}\label{eq:FkGk}
\begin{split}
|n_k(x,t,y)\rangle &=  e^{-i(\lambda_k^+ +\lambda_k^{+,2}t +\lambda_k^{+,3}y)J} \left(\begin{array}{c} c_{0k}^+ \\ \sqrt{2}  |\vec{\nu}_{0k}\rangle  \\ (c_{0k}^+)^{-1} \end{array}\right)
= \left( \begin{array}{c} e^{z_k^+ -i\phi_k^+} \\ \sqrt{2}  |\vec{\nu}_{0k}\rangle \\ e^{-z_k^+ +i\phi_k^+} \end{array} \right), \\
\langle m_k(x,t,y)| &= \left( c_{0k}^-, \sqrt{2} \langle \eta_{0k}| , ( c_{0k}^-)^{-1} \right) |e^{i(\lambda_k^- +\lambda_k^{-,2}t +\lambda_k^{-,3}y)J} \\
&= \left( \begin{array}{c} e^{z_k^- +i\phi_k^-}, \; \langle \vec{\eta}_{0k}| \sqrt{2}  , \;  e^{-z_k^--i\phi_k^-} \end{array} \right),
\end{split}
\end{equation}
where
\begin{equation}\label{eq:njxt}
\begin{aligned}
z_k^\pm  &= \nu_k^\pm \left( x+2\mu_k^\pm t + (3\mu_k^{\pm,2}-\nu_k^{\pm,2})y \right) + \xi_{0k}^\pm , &\quad \xi_{0k}^\pm &= \ln \left|c_{0k}^\pm \right| \\
\phi_k^\pm &=  \mu_k^\pm  x + (\mu_k^{\pm,2} -\nu_k^{\pm, 2}) t  + (\mu_k^{\pm,3} -3 \mu_k^{\pm} \nu_k^{\pm, 2})y + \phi_{0k}^\pm , & \quad \phi_{0k}^\pm &= \arg (c_{0k}^\pm ).
\end{aligned}
\end{equation}
and $|\vec{\nu}_{0k}\rangle $ and $\langle \vec{\eta}_{0k}| $ are constant $2n-1$-component polarization vectors normalized by $\langle \vec{\nu}_{0k}|\vec{\nu}_{0k}\rangle =1 $ and $\langle \vec{\eta}_{0k}|\vec{\eta}_{0k}\rangle =1 $. It is easy to check that
\begin{equation}\label{eq:nkSnk}\begin{aligned}
\langle n_k (x,t,y)| S_0|n_k(x,t,y)\rangle &= \langle n_{0k}| S_0|n_{0k}\rangle= 2 (1- \langle \nu_{0k}|s_0| \nu_{0k} \rangle ) =0, \\
\langle m_k (x,t,y)| S_0|m_k(x,t,y)\rangle &= \langle m_{0k}| S_0|m_{0k}\rangle= 2 (1- \langle \eta_{0k}|s_0 | \eta_{0k} \rangle ) =0,
\end{aligned}\end{equation}
so the normalization conditions on the vectors $| \nu_{0k} \rangle$ and $| \eta_{0k} \rangle$ ensures the relations (\ref{eq:m0s0m0}).
Then the one-soliton solution $\vec{q}_{\rm 1s}$ of both KSS  and mKdV is
\begin{equation}\label{eq:q1s}
\begin{split}
\vec{q}_{\rm 1s}(x,t,y;z_1,\phi_1) &=\frac{\sqrt{2}(\lambda_1^- - \lambda_1^+)}{\langle m_1(x,t,y) | n_1(x,t,y) \rangle} \left( e^{-z_1^- -i\phi_1^-} s_0|\vec{\nu}_{01}\rangle +  e^{z_1^+ -i\phi_1^+} |\vec{\eta}_{01}\rangle \right) , \\
\vec{p}_{\rm 1s}(x,t,y;z_1,\phi_1) &=-\frac{\sqrt{2}(\lambda_1^- - \lambda_1^+)}{\langle m_1(x,t,y) | n_1(x,t,y) \rangle} \left( e^{z_1^- +i\phi_1^-} |\vec{\nu}_{01}\rangle +  e^{-z_1^+ +i\phi_1^+} s_0 |\vec{\eta}_{01}\rangle \right) , \\
\langle m_1(x,t,y) | n_1(x,t,y)\rangle &=  e^{2z_1 -i\phi_1} + 2 \langle \vec{\eta}_{01}|\vec{\nu}_{01}\rangle +e^{-2z_1 +i\phi_1} , \\
\phi_1(x,t,y) &= \phi_1^+ -\phi_1^-, \qquad z_1(x,t,y) = \frac{1}{2}(z_1^+ + z_1^-),
\end{split}
\end{equation}

We end this subsection by calculating the basic expressions that we will need for deriving the soliton solution of Hirota-Ohta:
\begin{equation}\label{eq:uvw}\begin{split}
 (\vec{q}_{\rm 1s}^T \vec{p}_{\rm 1s}) &=  -\frac{2(\lambda_1^- - \lambda_1^+)^2 }{(\langle m_1 | n_1 \rangle)^2} \left( 2 + \langle \eta_{01} | \nu_{01} \rangle ( e^{2z_1 - i \phi_1} + e^{-2z_1 + i \phi_1}) \right), \\
(\vec{q}_{\rm 1s}^T s_0\vec{q}_{\rm 1s}) &=  \frac{2(\lambda_1^- - \lambda_1^+)^2 }{(\langle m_1 | n_1 \rangle)^2} \left(  e^{2z_1^+-2i \phi_1^+} +  e^{-2z_1^- -2i \phi_1^-} + 2\langle \eta_{01} | \nu_{01} \rangle  e^{z_1^+ - z_1^- - i (\phi_1^+ + \phi_1^-)}  \right), \\
(\vec{p}_{\rm 1s}^T s_0\vec{p}_{\rm 1s}) &=  \frac{2(\lambda_1^- - \lambda_1^+)^2 }{(\langle m_1 | n_1 \rangle)^2} \left(  e^{-2z_1^+ +2i \phi_1^+} + e^{2z_1^- +2i \phi_1^-} + 2\langle \eta_{01} | \nu_{01} \rangle  e^{-z_1^+ + z_1^- + i (\phi_1^+ + \phi_1^-)}  \right).
\end{split}\end{equation}

\subsection{The $2$-soliton solutions and soliton interactions }

We will derive the two-soliton solution by using an alternative version of the dressing method \cite{ZMNP, ZaMi, ZaSha1, ZaSha2}. Since we know the one-soliton solution and its dressing factor $u_1(x,t,y,\lambda)$ we can easily construct the FAS of the Lax operator $L_1$ whose potential is $Q_{\rm 1s}(x,t,y)$:
\begin{equation}\label{eq:chi1}\begin{aligned}
\chi_1^\pm(x,t,y,\lambda) & = u_1(x,t,y,\lambda) e^{-i (\lambda x + \lambda^2 t +\lambda^3 y))J} , \\
\hat{\chi}_1^\pm(x,t,y,\lambda) & = e^{i (\lambda x + \lambda^2 t +\lambda^3 y))J}  \hat{u}_1(x,t,y,\lambda) .
\end{aligned}\end{equation}
Now we will apply the dressing method again starting with the FAS $\chi_1^\pm(x,t,y,\lambda)$ and applying to them dressing factor $u_2(x,t,y,\lambda)$ having simple pole singularities at $\lambda_2^\pm$. This dressing factor will be:
\begin{equation}\label{eq:71.2}\begin{split}
&u_2(x,t,y,\lambda) = \openone + (c_2(\lambda)-1) \P_2(x,t,y) + (c_2^{-1}(\lambda)-1) \bar{\P}_2(x,t,y) ,\\
&c_2(\lambda) = \frac{\lambda -\lambda_2^+}{\lambda -\lambda_2^-}, \quad \P_2(x,t,y) = \frac{|\n_2 (x,t,y) \rangle \langle \m_2(x,t,y) |}{\langle \m_2(x,t,y)| \n_2(x,t,y) \rangle}, \quad \bar{\P}_2(x,t,y) = S_0 \P_2^T(x,t,y) S_0,  \\
&|\n_2(x,t,y) \rangle = u_1(x,t,y,\lambda_2^+) |n_2 \rangle ,  \quad \langle \m_2 | =  \langle m_2 | \hat{u} (x,t,y,\lambda_2^-)
\end{split}\end{equation}
where
\begin{equation}\label{eq:71.4}\begin{aligned}
|n_2 \rangle &= e^{-i(\lambda_2^+ x +\lambda_2^{+,2} t  +\lambda_2^{+,3}y)J} \left(\begin{array}{c} c_{02}^+ \\ \sqrt{2} |\nu_{02} \rangle \\ \hat{c}_{02}^+  \end{array}\right) =\left(\begin{array}{c} e^{z_2^+ - i \phi_2^+} \\ \sqrt{2} |\nu_{02} \rangle \\ e^{-z_2^+ + i \phi_2^+}  \end{array}\right), \\
\langle m_2 | &= (c_{02}^- , \sqrt{2} \langle \eta_{02}| , \hat{c}_{02}^-) e^{i (\lambda_2^-x + \lambda_2^{-,2}t +\lambda_2^{-,3} y)J} = (  e^{z_2^- + i \phi_2^-}, \sqrt{2} \langle \eta_{02}|  , e^{-z_2^- - i \phi_2^-}).
\end{aligned}\end{equation}
Thus we obtain:
\begin{equation}\label{eq:n2bold}\begin{aligned}
|\n_2 \rangle &= |n_2 \rangle + \frac{\lambda_1^- -\lambda_1^+}{\lambda_2^+ -\lambda_1^-} \frac{\langle m_1 | n_2 \rangle }{ \langle m_1 | n_1 \rangle}  |n_1 \rangle    +  \frac{\lambda_1^+ -\lambda_1^-}{\lambda_2^+ -\lambda_1^+} \frac{\langle n_1 |S_0| n_2 \rangle }{ \langle m_1 | n_1 \rangle} S_0|m_1 \rangle, \\
\langle \m_2 | &=  \langle m_2 | + \langle n_1 |S_0 \frac{\lambda_1^- -\lambda_1^+}{\lambda_2^- -\lambda_1^-} \frac{\langle m_2|S_0 | m_1 \rangle }{ \langle m_1 | n_1 \rangle}   +  \langle m_1 | \frac{\lambda_1^- -\lambda_1^+}{\lambda_2^- -\lambda_1^-} \frac{\langle m_2 | n_1 \rangle }{ \langle m_1 | n_1 \rangle}.
\end{aligned}\end{equation}
We will also need the scalar product:
\begin{equation}\label{eq:71.7}\begin{split}
\langle \m_2 |\n_2 \rangle &= \langle m_2 |n_2 \rangle + \frac{(\lambda_1^+ -\lambda_1^-)(\lambda_2^+ -\lambda_1^-)}{(\lambda_2^- -\lambda_1^-)(\lambda_2^+ -\lambda_1^+)}
\frac{\langle m_2 | n_1 \rangle \langle m_1 | n_2 \rangle}{ \langle m_1 | n_1 \rangle} \\
&- \frac{(\lambda_1^+ -\lambda_1^-)(\lambda_2^+ -\lambda_2^-)}{(\lambda_2^- -\lambda_1^-)(\lambda_2^+ -\lambda_1^+)} \frac{\langle n_1|S_0 | n_2 \rangle \langle m_2|S_0 | m_1 \rangle}{ \langle m_1 | n_1 \rangle}
\end{split}\end{equation}
The 2-soliton solution is obtained from:
\begin{equation}\label{eq:Q2s}\begin{split}
 Q_{\rm 2s}(x,t) = (\lambda_2^- -\lambda_2^+) [J, \P_2 - \bar{\P}_2 ] + (\lambda_1^- -\lambda_1^+) [J, P_1 - \bar{P}_1 ].
\end{split}\end{equation}


Let us now analyze the soliton interactions for the generic KSS. To this end we need to: i) fix up frame of reference  in such a way that the center-of-mass of the second soliton is fixed up at the origin; ii) calculate the limits of the dressing factor $u_1(x,t,\lambda)$ for $t \to \pm \infty$ and/or $y \to \pm \infty$ in this frame. Let us assume that we have fixed up the $k$-th soliton and let us introduce some more transparent notations which we will use also for analyzing the $N$-soliton interactions in the frame of reference of the $k$-th soliton:
\begin{equation}\label{eq:notk}\begin{aligned}
z_k &= \frac{1}{2} (z_k^+ + z_k^-)= \nu_k (x + v_k t + w_k y) + \xi_{0k}, &\quad
v_k &= \frac{\mu_k^+ \nu_k^+ + \mu_k^- \nu_k^-}{\nu_k}, \\
w_k &=  \frac{3\mu_k^{+,2} +3\mu_k^{-,2} -\nu_k^{+,2}-\nu_k^{-,2}  }{2\nu_k}   ,
& \quad \nu_k &= \frac{1}{2} (\nu_k^+ + \nu_k^-),
\end{aligned}\end{equation}
and $\xi_{0k} = \frac{1}{2} (\xi_{0k}^+ + \xi_{0k}^-)$.
Next we fix $z_k = Z_{0k} =\const.$ Then in this frame of reference
\begin{equation}\label{eq:x-fix}\begin{split}
 x = \frac{1}{\nu_k} (Z_{0k} - \xi_{0k}) - v_k t - w_k y.
\end{split}\end{equation}
Therefore, calculating $z_1$ in this frame of reference we get:
\begin{equation}\label{eq:tauk}\begin{split}
 \tau_{1,k} &= z_1|_{z_k= \rm fix} = \nu_1 (v_1 -v_k)t +  \nu_1 (w_1 -w_k)y + \tau_{0;1k}, \\ \tau_{0;1k} &= \frac{\nu_1}{\nu_k} (Z_{0k} -\xi_{0k}) +\xi_{01}.
\end{split}\end{equation}

\begin{remark}\label{rem:4}
Let us analyze the interactions of the KSS solitons. Then we need to evaluate the asymptotics of the dressing factor $u_1(x,t,y,\lambda)$ for $t\to \pm\infty$ at fixed $z_k$ and $y$. Obviously this could be done only for the generic case when all solitons have different velocities, i.e. $v_j \neq v_k$ for $j\neq k$. If we need to analyze the interactions of the mKdV solitons we will need to evaluate the asymptotics of the dressing factors $u_1(x,t,y,\lambda)$ for $y\to \pm\infty$ at fixed $z_k$ and $t$. Again this could be done only for the generic case when all solitons have different velocities, i.e. $w_j \neq w_k$ for $j\neq k$.
\end{remark}

For the two-soliton case we take $k=2$ and obtain:
\begin{equation}\label{eq:taukpm}\begin{aligned}
\lim_{\tau_{12}\to \infty} u_1(x,t,y,\lambda) &= u_1^+(\lambda), &\quad \lim_{\tau_{12}\to -\infty} u_1(x,t,y,\lambda) &= u_1^-(\lambda), &\quad \mbox{for} &\quad  v_1 >v_2, \\
\lim_{\tau_{12}\to \infty} u_1(x,t,y,\lambda) &= u_1^-(\lambda), &\quad \lim_{\tau_{12}\to -\infty} u_1(x,t,y,\lambda) &= u_1^+(\lambda), &\quad \mbox{for} &\quad  v_1 <v_2,
\end{aligned}\end{equation}
where
\begin{equation}\label{eq:u1as}\begin{split}
u_1^+(\lambda) &= \exp (\ln c_1(\lambda) J) = \diag (c_1(\lambda), \openone, c_1^{-1}(\lambda)), \\ u_1^-(\lambda) &= \exp (-\ln c_1(\lambda) J) = \diag (c_1^{-1}(\lambda), \openone , c_1(\lambda)),
\end{split}\end{equation}

Note that the asymptotic of $u_1(x,t,y,\lambda)$ depends on $\lambda$ and on the sign of $v_1 - v_2$. As a result for the asymptotics of $\langle \m_2 |$ and $\n_2 \rangle$ we find:
\begin{equation}\label{eq:limn28}\begin{aligned}
&\lim_{\tau_{12}\to \infty}|\n_2 \rangle = e^{(\ln c_1(\lambda_2^+) J)}|\n_2 \rangle \quad & \quad &\lim_{\tau_{12}\to -\infty}|\n_2 \rangle = e^{(-\ln c_1(\lambda_2^+) J)}|\n_2 \rangle , \\
&z_{2+, \rm as}^+ = z_2^+ + \ln \left| \frac{\lambda_2^+ -\lambda_1^+}{\lambda_2^+ -\lambda_1^-}\right|,  \quad & \quad  &z_{2-, \rm as}^+ = z_2^+ - \ln \left| \frac{\lambda_2^+ -\lambda_1^+}{\lambda_2^+ -\lambda_1^-}\right|,\\
&\phi_{2+, \rm as}^+ = \phi_2^+ - \arg \left( \frac{\lambda_2^+ -\lambda_1^+}{\lambda_2^+ -\lambda_1^-}\right), \quad & \quad &\phi_{2-, \rm as}^+ = \phi_2^+ + \arg \left( \frac{\lambda_2^+ -\lambda_1^+}{\lambda_2^+ -\lambda_1^-}\right)  \\
\end{aligned}\end{equation}
and
\begin{equation}\label{eq:limm28}\begin{aligned}
&\lim_{\tau_{12}\to \infty} \langle \m_2| =  \langle m_2|e^{(\ln c_1(\lambda_2^-)J}),  \quad &  \quad &\lim_{\tau_{12}\to -\infty} \langle \m_2| =  \langle m_2|e^{(-\ln c_1(\lambda_2^-) J)}, \\
&z_{2+, \rm as}^- = z_2^- - \ln \left| \frac{\lambda_2^- -\lambda_1^+}{\lambda_2^- -\lambda_1^-}\right|,  \quad & \quad &z_{2-, \rm as}^- = z_2^- + \ln \left| \frac{\lambda_2^- -\lambda_1^+}{\lambda_2^- -\lambda_1^-}\right|, \\
&\phi_{2+, \rm as}^- = \phi_2^- - \arg \left( \frac{\lambda_2^- -\lambda_1^+}{\lambda_2^- -\lambda_1^-}\right),
\quad & \quad &\phi_{2-, \rm as}^- = \phi_2^- + \arg \left( \frac{\lambda_2^- -\lambda_1^+}{\lambda_2^- -\lambda_1^-}\right)
\end{aligned}\end{equation}
As a consequence the shift of the center-of-mass of the second soliton is provided by:
\begin{equation}\label{eq:limzpm}\begin{aligned}
 z_{2,\pm \rm as} &=\lim_{\tilde{x}\to \pm \infty} \frac{1}{2}(z_2^+ + z_2^-)  = z \pm \xi_{0;2,1},  &  \xi_{0;2,1} &=  \frac{1}{2} \ln \left|  c_1(\lambda_2^+)c_1(\lambda_2^-)  \right|, \\
\phi _{2,\pm \rm as} &=\lim_{\tilde{x}\to \infty} (-\phi_2^+ + \phi_2^-) =(-\phi_2^+ + \phi_2^-) \pm \varphi_{0;2,1} , & \varphi_{0;2,1} &= \arg  \left(  c_1(\lambda_2^+)c_1(\lambda_2^-)  \right),
\end{aligned}\end{equation}

\subsection{$N$-soliton interaction for the KSS}

It will be more convenient for us to construct the $N$-soliton solution
by applying  $N$ times subsequently the above procedure with the result:
\begin{equation}\label{eq:25.1ns}\begin{split}
u_{\rm Ns}(x,t,y,\lambda) = u_N(x,t,y,\lambda) \cdots u_1(x,t,y,\lambda)
\end{split}\end{equation}
where the individual dressing factors take the form (see \cite{Basic, Mon1}):
\begin{equation}\label{eq:19.0}\begin{aligned}
u_s(x,t,y,\lambda) &= \openone + (c_s(\lambda) -1) \P_s(x,t,y) + (c_s^{-1}(\lambda) -1) \bar{\P}_s(x,t,y) , &\quad \P_s = \frac{ |\n_s \rangle \langle \m_s|} {\langle \m_s| \n_s \rangle}, \\
|\n_s \rangle & = u_{s-1}(x,t,y,\lambda_s^+) \cdots u_{1}(x,t,y,\lambda_s^+) |n_{s}\rangle , \\
\langle \m_s| &= \langle m_{s}| \hat{u} _{s-1}(x,t,y,\lambda_s^-) \cdots \hat{u}_{1}(x,t,y,\lambda_s^-) ,
\end{aligned}\end{equation}
where the vectors $|n_s \rangle$ and $ \langle m_s|$ are given in (\ref{eq:FkGk}).

This leads to the following results for the asymptotics of the dressing factors:
\begin{equation}\label{eq:u1-as}\begin{split}
\lim_{\tau_{sN}\to \infty} u_s(x,t) \equiv u_s^+(\lambda)  = e^{ J \ln c_s(\lambda) }, \qquad
\lim_{\tau_{sN} \to -\infty}u_s(x,t)  \equiv u_s^-(\lambda) = e^{- J \ln c_s(\lambda) },
\end{split}\end{equation}
if $v_s > v_N$ and
\begin{equation}\label{eq:u1-ass}\begin{split}
\lim_{\tau_{sN}\to \infty} u_s(x,t)   = e^{ -J \ln c_s(\lambda) }, \qquad
\lim_{\tau_{sN} \to -\infty} u_s(x,t)  = e^{ J \ln c_s(\lambda) },
\end{split}\end{equation}
for $v_s < v_N$.

\begin{remark}\label{rem:3}
Note that all $u_s^\pm(\lambda)$ depend only on $\lambda$, and are given by
diagonal matrices;  hence, they commute.

\end{remark}

Finally we evaluate the limits of $\P_N(x,t)$ with the result:
\begin{equation}\label{eq:P-N}\begin{aligned}
&\lim_{\tau_{sN} \to \infty} P_N(x,t) = \frac{|\n_N^+\rangle \langle m_N^+|}{\langle m_N^+|  n_N^+\rangle} , &\qquad
& \lim_{\tau_{sN} \to -\infty} P_s(x,t) = \frac{|\n_N^-\rangle \langle m_N^-|}{\langle m_N^-|  n_N^-\rangle} , \\
& |\n_N^+\rangle = \prod_{s\in \sigma_\mu^+}^{} (u_s^+(\lambda_N^+))^2 |n_N\rangle , & \qquad
&|\n_N^-\rangle = \prod_{s\in \sigma_\mu^-}^{} (u_s^+(\lambda_N^+))^{-2} |n_N\rangle ,
\end{aligned}\end{equation}
where $\sigma^+$ (resp.  $\sigma^-$) are the sets of indices $1\leq s \leq N-1$ for which $v_s >v_N$ (resp.  $v_s <v_N$). Thus we find:
\begin{equation}\label{eq:z-Np}\begin{aligned}
\lim_{\tau_{sN}\to \infty} Q_{\rm Ns}(x,t) &= Q_{\rm 1s}(z_N+ \zeta_N^+ , \phi_N +\varphi_N^+), \\ \lim_{\tau_{sN}\to -\infty} Q_{\rm Ns}(x,t) &= Q_{\rm 1s}(z_N -\zeta_N^-, \phi_N - \varphi_N^-),
\end{aligned}\end{equation}
where
\begin{equation}\label{eq:z-No1}\begin{aligned}
\zeta_N^+ & =  \frac{1}{2}\sum_{s\in \sigma_\mu^+}^{} \ln \left| c_s(\lambda_N^+) c_s(\lambda_N^-) \right|, &\qquad
\zeta_N^- & =  2\sum_{s\in \sigma_\mu^-}^{} \ln \left|  c_s(\lambda_N^+) c_s(\lambda_N^-) \right| \\
\varphi_N^+ & =  2\sum_{s\in \sigma_\mu^+}^{} \arg  \left(  c_s(\lambda_N^+) c_s(\lambda_N^-) \right), &\qquad
\varphi_N^- & =  2\sum_{s\in \sigma_\mu^-}^{} \arg  \left(  c_s(\lambda_N^+) c_s(\lambda_N^-) \right).
\end{aligned}\end{equation}
The above formulae show that the $N$-soliton interactions of the KSS (or of the {\bf BD.I}-type MNLS are like for the scalar NLS case. First of all they are {\em purely elastic}, i.e. no creation or annihilation of solitons is possible. The only effect of the interaction is  the shifts of relative center-of-mass coordinates and of the phases, while the polarization vectors $\vec{\nu}_{s0}$ are preserved.

Similar analysis can be applied also for the $N$-soliton interactions of mKdV solitons. In this case we will need to calculate the limits of $u_s(x,t,y,\lambda)$ for $y\to \pm \infty$ with fixed $z_k$ and $t$. The results will be quite analogous as for the KSS soliton interactions, see
Remark~\ref{rem:4} above.

We end this Section by stating the fact that this approach allows us to recover explicitly also the FAS $\chi^\pm (x,t,y,\lambda)$ of the Lax triple. Indeed, deriving the $N$-soliton solutions we started with the trivial potential $Q_0(x,t,y) =0$ and
\[ \chi_0^\pm (x,t,y,\lambda)= \exp( -i\lambda (x+\lambda t +\lambda^2y)J). \]
Therefore the FAS are given by:
\begin{equation}\label{eq:chiN}\begin{split}
 \chi^\pm (x,t,y,\lambda) = u_N(x,t,y,\lambda) \chi_0^\pm (x,t,y,\lambda)
\end{split}\end{equation}
This allows us to calculate also the Gauss factors of the scattering matrix $S^\pm (y,t,\lambda)$, $T^\pm (y,t,\lambda)$ and $D^\pm (\lambda)$.

The next step could be to derive the expansions of $Q_{\rm Ns}(x,t,y)$ and its variation $\ad_J^{-1} \delta Q_{\rm Ns}(x,t,y)$ over the `squared' solutions of the Lax operator, see e.g. \cite{VSG2, Basic, EPJPI}. The corresponding expansion coefficients corresponding to  the continuous spectrum of $L$ will all be vanishing. Thus, restricting ourselves to the class of $N$-soliton solutions we in fact obtain a dynamical system with finite number of degrees of freedom which must be superintegrable. These systems will have also several compatible Hamiltonian formulations.  These aspects will be published elsewhere.

\section{ Multi-Soliton Solutions of the Hirota--Ohta system}

Here we will show that our results can be extended to the
Hirota-Ohta system \cite{HiOh1, HiOh2, Adler}.

The Hirota--Ohta system
\begin{equation}\label{eq:01}\begin{split}
 4w_{y} &=w_{xxx}+6ww_{x}+3\partial _{x}^{-1}w_{tt}-24(uv)_{x}, \\
 -2u_{y} &=u_{xxx}-3u_{xt}+3wu_{x}-3(\partial _{x}^{-1}w_{t})u, \\
 -2v_{y} &=v_{xxx}+3v_{xt}+3wv_{x}+3(\partial _{x}^{-1}w_{t})v,
\end{split}\end{equation}
follows from the Lax representation
\begin{equation}
\psi _{t}=\psi _{xx}+w\psi +2u\varphi ,\qquad -\varphi _{t}=\varphi
_{xx}+w\varphi +2v\psi ;  \label{1}
\end{equation}%
\begin{equation}
\psi _{y}=\psi _{xxx}+\frac{3}{2}w\psi _{x}+b\psi +3u_{x}\varphi ,\qquad %
\varphi _{y}=\varphi _{xxx}+\frac{3}{2}w\varphi _{x}+n\varphi +3v_{x}\psi ,
\label{2}
\end{equation}%
where%
\begin{equation*}
b=\frac{3}{4}(w_{x}+\partial _{x}^{-1}w_{t}),\qquad  n=\frac{3}{4}%
(w_{x}-\partial _{x}^{-1}w_{t}).
\end{equation*}

The compatibility conditions of the equations (\ref{1}) containing a spectral parameter $\lambda$:
\begin{equation}
\lambda \psi =\psi _{x}+\overset{2n-1}{\underset{k=1}{\sum }}\hat{p}_{k}\partial
_{x}^{-1}(\hat{q}_{k}\psi +\hat{p}_{k}\varphi ), \qquad -\lambda \varphi =\varphi _{x}+%
\overset{2n-1}{\underset{k=1}{\sum }}\hat{q}_{k}\partial _{x}^{-1}(\hat{q}_{k}\psi
+\hat{p}_{k}\varphi )  \label{3}
\end{equation}%
lead to the Kulish--Sklyanin system (\ref{eq:ks2}).

\begin{remark}\label{rem:ksi}
The variables $\hat{q}_{k}$ and $\hat{p}_{k}$ are linearly related, so that
\[ \sum_{k=1}^{2n-1} \hat{q}_{k}^2 = (\vec{q}^T s_0 \vec{q}), \qquad \sum_{k=1}^{2n-1} \hat{p}_{k}^2 = -2(\vec{p}^T s_0 \vec{p}), \qquad \sum_{k=1}^{2n-1} \hat{q}_{k} \hat{p}_{k} = -2(\vec{q}^T \vec{p}). \]
Another small but important difference between  (\ref{eq:ks2}) and (\ref {KS}) is transformation of the time variable  $t$ from real to purely imaginary: $t\rightarrow  -it$.
\begin{equation}
\hat{p}_{k,t}=\hat{p}_{k,xx}+w\hat{p}_{k}-2u\hat{q}_{k},\qquad -\hat{q}_{k,t}=\hat{q}_{k,xx}+w\hat{q}_{k}-2v\hat{p}_{k},
\label{KS}
\end{equation}%
and its higher commuting flow (cf. (\ref{eq:ks3}))%
\begin{equation}\label{eq:4.5a}
\hat{p}_{k,y}=\hat{p}_{k,xxx}+\frac{3}{2}w\hat{p}_{k,x}+b\hat{p}_{k}-3\hat{q}_{k}u_{x}, \qquad %
\hat{q}_{k,y}=\hat{q}_{k,xxx}+\frac{3}{2}w\hat{q}_{k,x}+n\hat{q}_{k}-3\hat{p}_{k}v_{x},
\end{equation}%
where the reduction of the Hirota--Ohta system is determined by the
constraints%
\begin{equation}
u=\frac{1}{2}\overset{2n-1}{\underset{k=1}{\sum }}\hat{p}_{k}^{2},\qquad  w=2%
\overset{2n-1}{\underset{k=1}{\sum }}\hat{p}_{k}\hat{q}_{k},\qquad v=\frac{1}{2}\overset{%
2n-1}{\underset{k=1}{\sum }}\hat{q}_{k}^{2},  \label{sums}
\end{equation}%
\begin{equation*}
b=3\overset{2n-1}{\underset{k=1}{\sum }}\hat{q}_{k}\hat{p}_{k,x},\text{ \ }n=3\overset{2n-1}{%
\underset{k=1}{\sum }}\hat{p}_{k}\hat{q}_{k,x}.
\end{equation*}
\end{remark}

Since we already constructed multi-soliton solutions for both above
two-dimensional commuting systems (\ref{KS}), a corresponding multi-soliton
solution of the Hirota--Ohta system is given by (\ref{sums}). {\small The
explicit expressions for the right hand sides of the above equations are in
eq. (\ref{eq:uvw}) above. Since we know the above relations we are able to
calculate a class of solutions to Hirota-Ohta model and analyze their
interactions.}

The above Lax triad (\ref{1}), (\ref{2}), (\ref{3}) is
written in scalar form for two unknown function $\psi $ and $\varphi $ only.
This Lax representation is very convenient for description of higher
commuting flows. However, Lax representation written in a matrix form is
more suitable for construction of multi-soliton solutions. Indeed, let us
introduce new notations:
\begin{equation}\label{eq:}\begin{split}
 \vec{X} &= (\psi, \chi_1, \chi_2, \dots , \chi_{2n-1}, \varphi )^T, \\
 \vec{Q} &= (\hat{q}_1, \hat{q}_2 , \dots, \hat{q}_{2n-1})^T, \qquad  \vec{P} = (\hat{p}_1, \hat{p}_2 , \dots, \hat{p}_{2n-1})^T.
\end{split}\end{equation}
 with which we can rewrite the scalar Lax representation in matrix form:

\begin{equation}
\chi _{k,x}=\hat{q}_{k}\psi +\hat{p}_{k}\varphi .  \label{uno}
\end{equation}%
Then (\ref{3}) becomes%
\begin{equation}
\psi _{x}=\lambda \psi -\overset{2n-1}{\underset{m=1}{\sum }}\hat{p}_{m}\chi _{m},%
\text{ \ }\varphi _{x}=-\lambda \varphi -\overset{2n-1}{\underset{m=1}{\sum }}%
\hat{q}_{m}\chi _{m}.  \label{due}
\end{equation}%
This means, altogether that the above equations can be written in matrix form
known in the theory of two-dimensional integrable systems%
\begin{equation}\label{eq:Lx}\begin{split}
 \frac{\partial \vec{X}}{\partial x } = \left(\begin{array}{ccc} \lambda & -\vec{P}^T & 0 \\
 \vec{Q} & 0 & \vec{P} \\ 0 & -\vec{Q}^T & - \lambda  \end{array}\right) \vec{X},
\end{split}\end{equation}

Then the first equation in the 'scalar' Lax representation (\ref{1})%
\begin{equation*}
\psi _{t}=\psi _{xx}+w\psi +2u\varphi ,\text{ \ }-\varphi _{t}=\varphi
_{xx}+w\varphi +2v\psi ;
\end{equation*}%
becomes%
\begin{equation}\label{eq:Lt}\begin{split}
\frac{\partial \vec{X}}{ \partial t} = \left(\begin{array}{ccc} \lambda^2 + \frac{1}{2} w & -\lambda \vec{P}^T - \vec{P}^T_x & 0 \\ \lambda \vec{Q}- \vec{Q}_x & \vec{P} \vec{Q}^T - \vec{Q} \vec{P}^T & \lambda \vec{P} + \vec{P}_x \\ 0 & -\lambda \vec{Q}^T + \vec{Q}^T_x & -\lambda^2 - \frac{1}{2} w  \end{array}\right)
\end{split}\end{equation}

\begin{remark}\label{rem:ort}
 Note that  the $3\times 3$ block matrices in the right hand sides of equations (\ref{eq:Lx}) and  (\ref{eq:Lt}) are orthogonal in the sense that they are anti-symmetric with respect to the second diagonal. The equivalence between (\ref{eq:Lax}) and (\ref{eq:Lx}) means that $q_k$ and $p_k$ are linearly related to $\hat{q}_k$ and $\hat{p}_k$, see Remark~\ref{rem:ksi}. Equation (\ref{eq:4.5a}) can also be written in matrix form, but this requires some cumbersome calculations which we will omit.
\end{remark}

\section{Discussion and conclusions}

In short we have established that the Lax triad $L$, $M_2$ and $M_3$ is equivalent to the scalar Lax representation (\ref{1}), (\ref{2}) and (\ref{3}).

 A similar procedure which allowed you to derive solutions of KP from  solutions of NLS  and a higher integral of motion
was proposed in \cite{DubMat1}. 

\section*{Acknowledgements}
We are grateful to Professor M. Pavlov for numerous useful discussions and suggestions.
The research of V. B. Matveev and Nianhua  Li was funded by RFBR and NSFC, project
number 21-51-53017. V. S. Gerdjikov has been supported in part by the Bulgarian Science Foundation,
contract KP-06N42-2. The work of A. O. Smirnov has been supported by the Ministry of Science and
Higher Education of the Russian Federation (grant agreement No. FSRF-2020-0004).

\end{document}